\documentclass{epsconf}

\usepackage{amsmath,graphicx,subfigure,amssymb,amsfonts,bm}

\title{Stable spatial Langmuir solitons as a model of long-lived atmospheric plasma structures}
\author{M. Dvornikov$^{1,2}$}
\institute{$^1$ Institute of Physics, University of S\~{a}o Paulo, S\~{a}o Paulo, Brazil \\
$^2$ Pushkov Institute of Terrestrial Magnetism, Ionosphere
and Radiowave Propagation (IZMIRAN), Troitsk, Moscow, Russia}

\begin{document}
\maketitle

A long-lived glowing structure in the atmosphere, sometimes called a \textit{ball lightning} (BL), attracts attention of the scientific community for almost two centuries since the pioneering work by Arago~\cite{Ara37}. Nevertheless, despite numerous theoretical and experimental efforts aimed to understand this phenomenon (see, e.g.,~\cite{Byc10}), physical processes underlying BL are still unclear.

BL has the following remarkable properties. According to the majority of eyewitness reports BL appears during a thunderstorm. Thus it should be associated with atmospheric electricity and have electrodynamic or plasma nature. BL is a glowing structure with the size of several centimeters. The lifetime of such an object can be up to several minutes. Taking into account the average lifetime of unstructured plasma and difficulties of the plasma confinement in a limited region of space, one can conclude that BL should have rather complicated internal structure. BL is reported to pass through microscopic cracks, with the size being much less than the visible dimensions of BL, in dielectric materials, e.g., glass, without further damaging them. It was even reported that BL burned microscopic holes and tracks while it was in contact with external materials. It means that BL should have a small, dense and hot core surrounded by a colder and dilute halo. The energy content of BL, indirectly estimated by the damage created while it disappears, is in the range from kJ to
MJ. It means that, in some situations, an additional energy source, which is not of the electromagnetic origin, can be present in BL.

We developed the model of BL based on radial nonlinear quantum oscillations of charged particles in plasma~\cite{DvoDvo07}. Note that spherically symmetric plasma oscillations do not radiate electromagnetic waves. Thus such a plasma structure will be stable. Besides the stability, one can explain various properties of BL in frames of this model like the possibility of passing through tiny holes and even the existence of the internal energy source. However, quantum effects are unlikely to be essential at the initial stages of the BL evolution. For this reason in~\cite{Dvo11,Dvo13} we proposed the classical electrodynamics description of BL, which involves stable spatial Langmuir solitons. This approach seems to be applicable at the stages of the BL formation.

In general situation the evolution of the electric field $\bm{E}(\mathbf{r},t)$ in a Langmuir wave packet can be presented in the following way: $\bm{E} = e^{-\mathrm{i}\omega_p t}\mathbf{E}(\mathbf{r},t) + \text{c.c.}$, where $\mathbf{E}$ is the envelope of the electric field, $\omega_p = \sqrt{4 \pi n_0 e^2/m}$ is the Langmuir frequency for electrons, and $n_0$ is the unperturbed electron density. Our main goal is to study the evolution of  $\mathbf{E}$ for radially symmetric plasma oscillations.
It is known that the stability of a Langmuir wave packet against the collapse is owing to the presence of a certain nonlinearity. The nonlinear effects based on the electron-electron interaction and the quantum Bohm pressure were discovered in~\cite{SkoHaa80} to result in the formation of stable spatial Langmuir solitons in two and three dimensions. In the present work we shall study Langmuir solitons in the cases of (i) nonlocal electron-electron nonlinearity; and (ii) nonlinearity based on the interaction of an induced electric dipole moment (EDM) of ions with the oscillating electric field $\bm{E}$.

In~\cite{Dvo11} we studied spatial Langmuir solitons in plasma which involves nonlocal electron-electron nonlinearity. We obtained there the following nonlinear Schr\"odinger equation (NLSE) for the description of radially symmetric plasma oscillations:
\begin{equation}\label{NLSE3}
  \lambda \psi = 
  - \Delta_x \psi -
  \psi |\psi|^2 + \beta(d) \frac{\psi |\psi|^2}{x^2} -
  \frac{3}{2}
  \left[
    \frac{3}{2} \Delta_x |\psi|^2 +
    \frac{d-1}{x} \frac{\partial |\psi|^2}{\partial x}
  \right]\psi,
\end{equation}
where $\psi(x) = \exp \left( \mathrm{i} \lambda s - \tfrac{9}{4}|\Phi|^2 \right) \Phi$ is the effective wave function, $\Delta_x = \tfrac{\partial^2}{\partial x^2} + \tfrac{(d-1)}{x} \tfrac{\partial}{\partial x}$ is the radial part of the Laplace operator, $\lambda$ is the dimensionless frequency shift, $\beta(d) = (d-1)(4-d)/2$, and $d=2,3$ is the dimension of space. The dimensionless parameters $x$, $s$, and $\Phi$ in Eq.~\eqref{NLSE3} are $x = \sqrt{\tfrac{3}{2}} \tfrac{r}{r_\mathrm{D}}$, $s = \tfrac{9}{4} \omega_p t$, and
$\Phi = \tfrac{|\mathbf{E}|}{\sqrt{18 \pi n_0 T_e}}$, where $r=|\mathbf{r}|$, $r_\mathrm{D} = \sqrt{T_{e}/4\pi e^{2}n_{0}}$ is the Debye length, and $T_e$ is the electron temperature.

\begin{figure}
  \centering
  \subfigure[]
  {\label{a}
  \includegraphics[scale=.051]{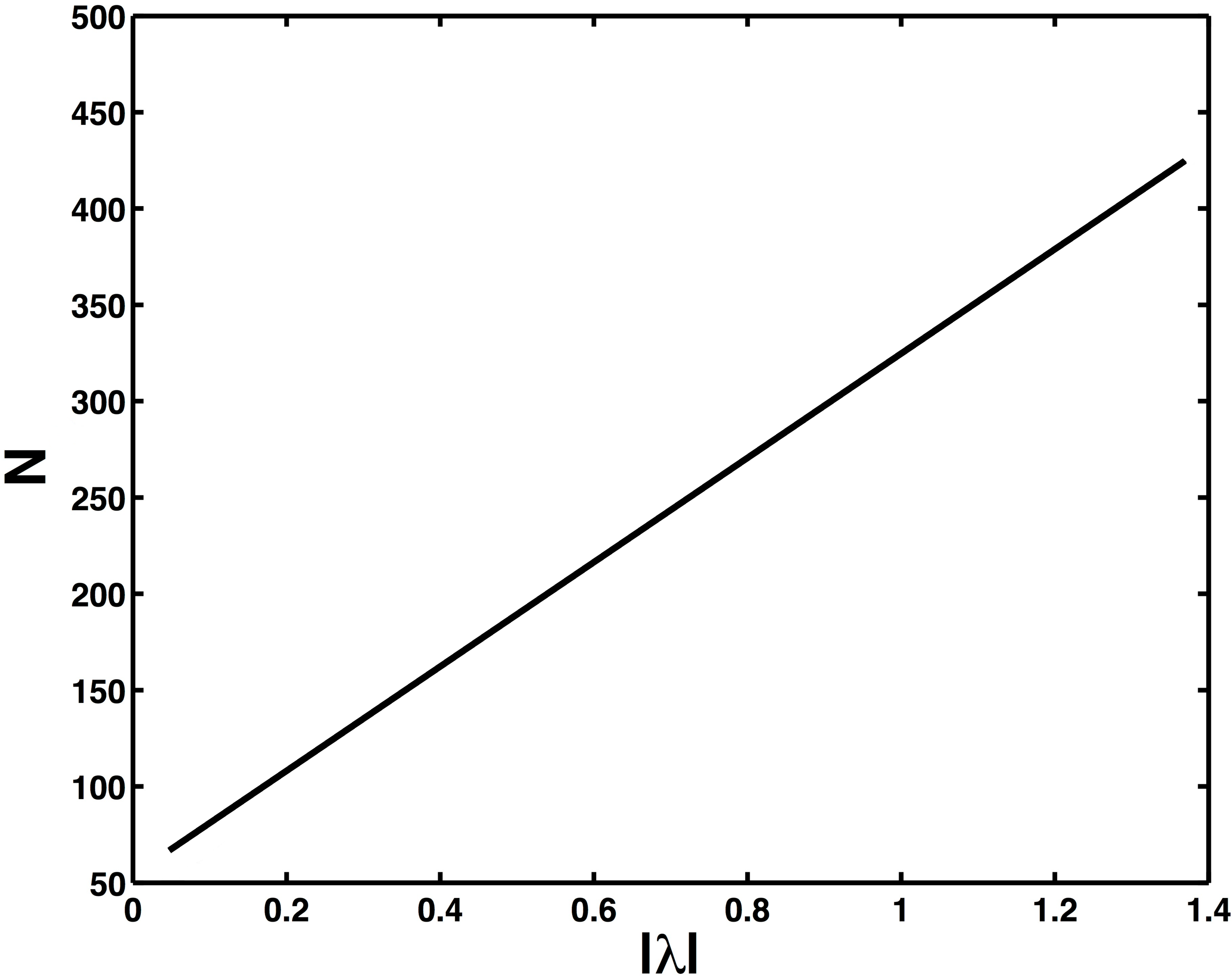}}
  \
  \subfigure[]
  {\label{b}
  \includegraphics[scale=.051]{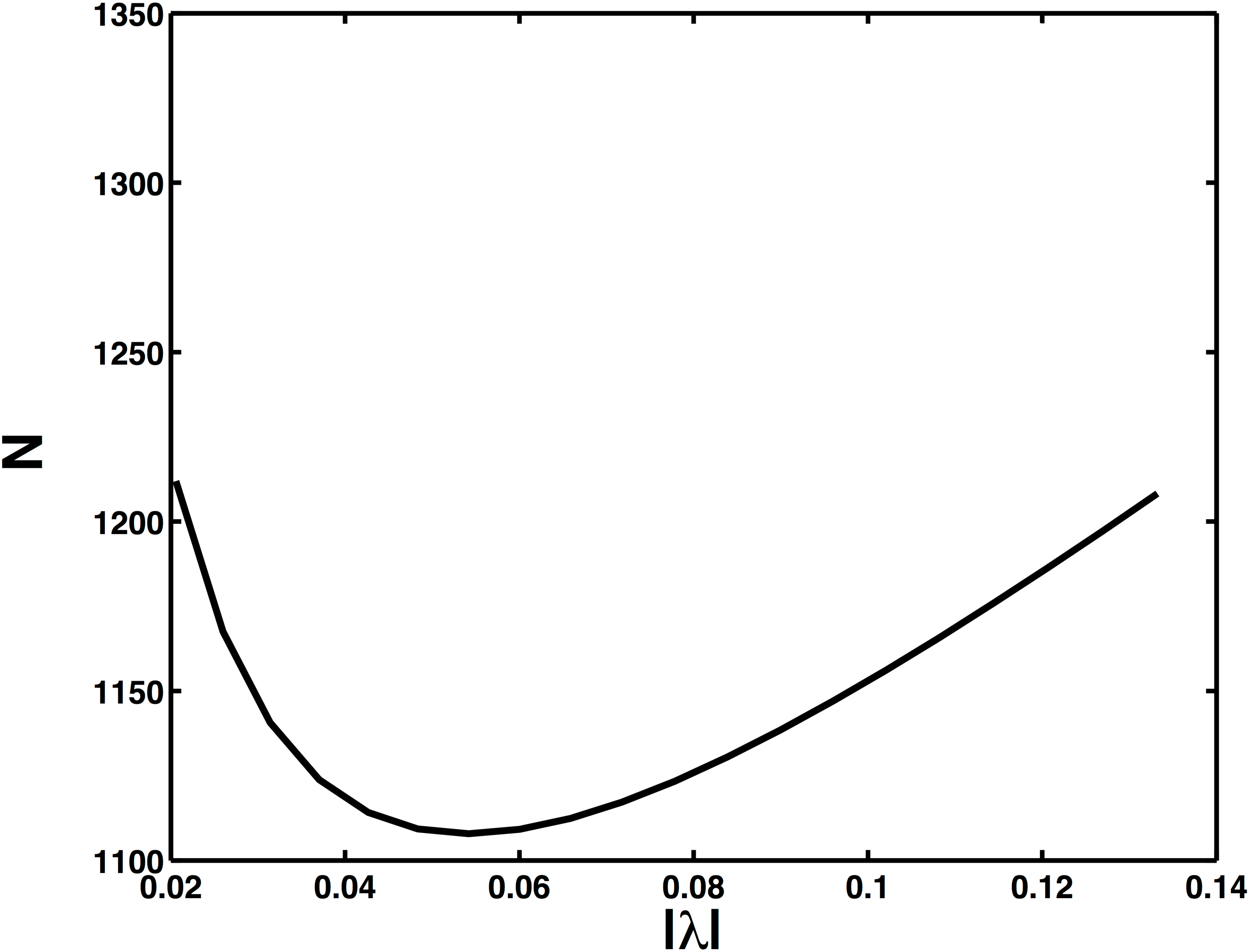}}
  \caption{Plasmon number versus the frequency shift for
  Eq.~\eqref{NLSE3} in
  (a) 2D case and (b) 3D case. 
  \label{ionosphBL}
  }
\end{figure}

Solutions of Eq.~\eqref{NLSE3} are found numerically. To analyze their stability in Fig.~\ref{ionosphBL} we show the plasmon number $N = \Omega_d \int_0^\infty \mathrm{d}x x^{d-1} |\psi|^2$ as a function of $\lambda$ in 2D, $d=2$, and 3D, $d=3$, cases. 
Here $\Omega_2 = 2\pi$ and $\Omega_3 = 4\pi$ is the solid angle in 2D and 3D spaces. Using the Vakhitov-Kolokolov criterion (VKC) we get that the solitons are stable in 2D case, since $\tfrac{\partial N}{\partial |\lambda|} > 0$, whereas in 3D space, stable and unstable, $\tfrac{\partial N}{\partial |\lambda|} < 0$, solutions coexist.

Another nonlinearity which arrests the collapse of Langmuir waves and results in the formation of stable spatial solitons is the interaction of induced EDM of ions with a rapidly oscillating electric field inside the plasmoid~\cite{Dvo13}. In this situation ions should be diatomic. The detailed analysis shows~\cite{Dvo13} that in this case the dynamics of a plasma soliton obeys the following cubic-quintic NLSE:
\begin{equation}\label{eq:NLSEpsi}
  \mathrm{i}\frac{\partial\psi}{\partial\tau} +
  \Delta_x \psi - \frac{d-1}{x^{2}}\psi +
  \left(
    |\psi|^{2}-|\psi|^{4}
  \right)
  \psi = 0,
\end{equation}
where  $\tau = \tfrac{15}{128\pi^{2}} \frac{T_{i}}{T_{e}}
\tfrac{\omega_{p}t}{(n_{0}\Delta\alpha)^{2}} $,
$x = \tfrac{\sqrt{5T_{i}/T_{e}}}{8\pi n_{0}\Delta\alpha}  \frac{r}{r_{\mathrm{D}}}$, $\psi = 4\Delta\alpha \sqrt{\frac{\pi n_{0}}{15T_{i}}}|\mathbf{E}|$, $T_i$ is the ion temperature, $\Delta\alpha = \alpha_{\parallel}-\alpha_{\perp}$, and $\alpha_{\parallel,\perp}$ are longitudinal and transversal polarizabilities of an ion.

\begin{figure}
  \centering
  \subfigure[]
  {\label{a}
  \includegraphics[width=7.35cm,height=5.2cm]{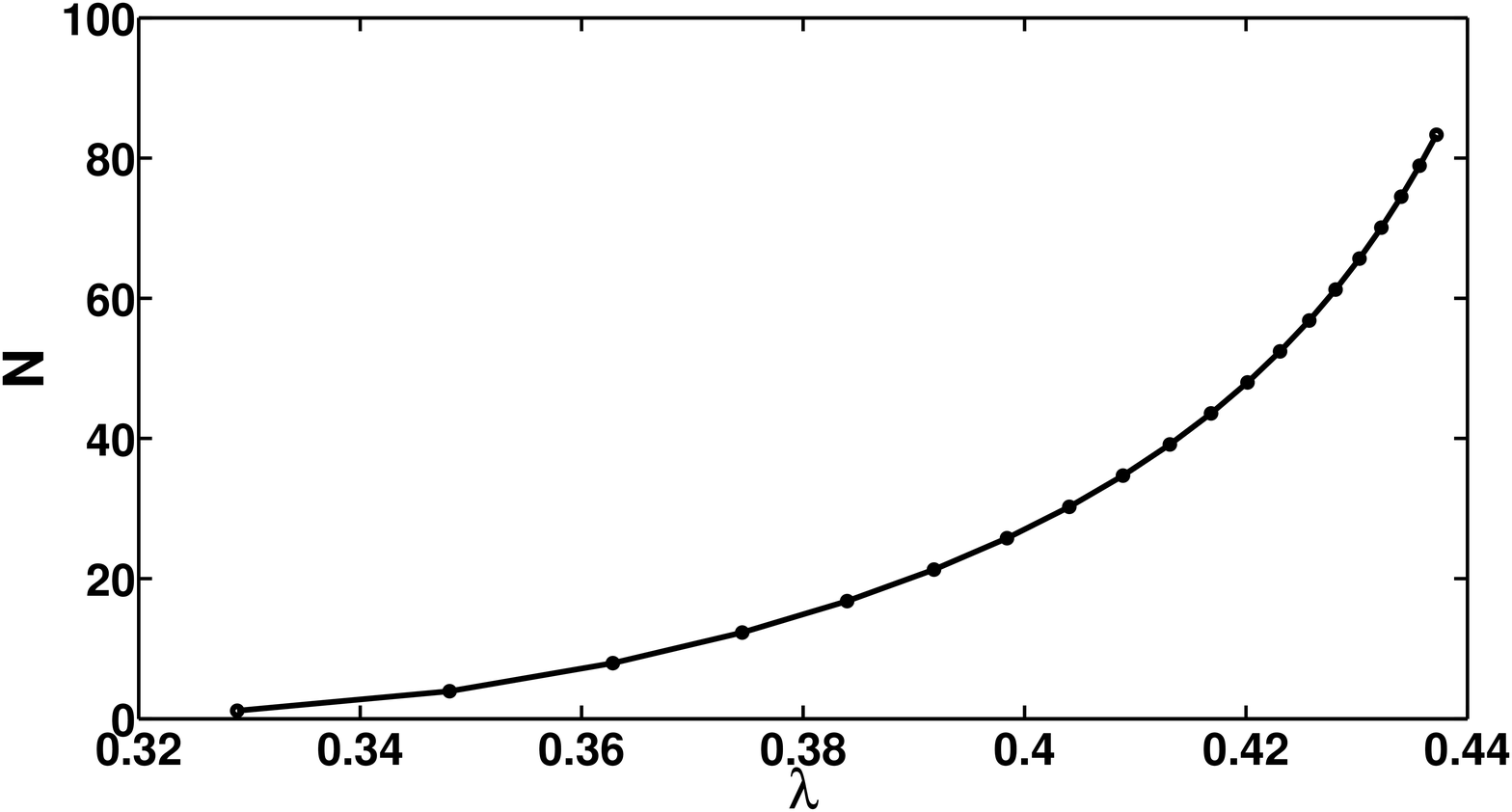}}
  \subfigure[]
  {\label{b}
  \includegraphics[width=7.35cm,height=5.2cm]{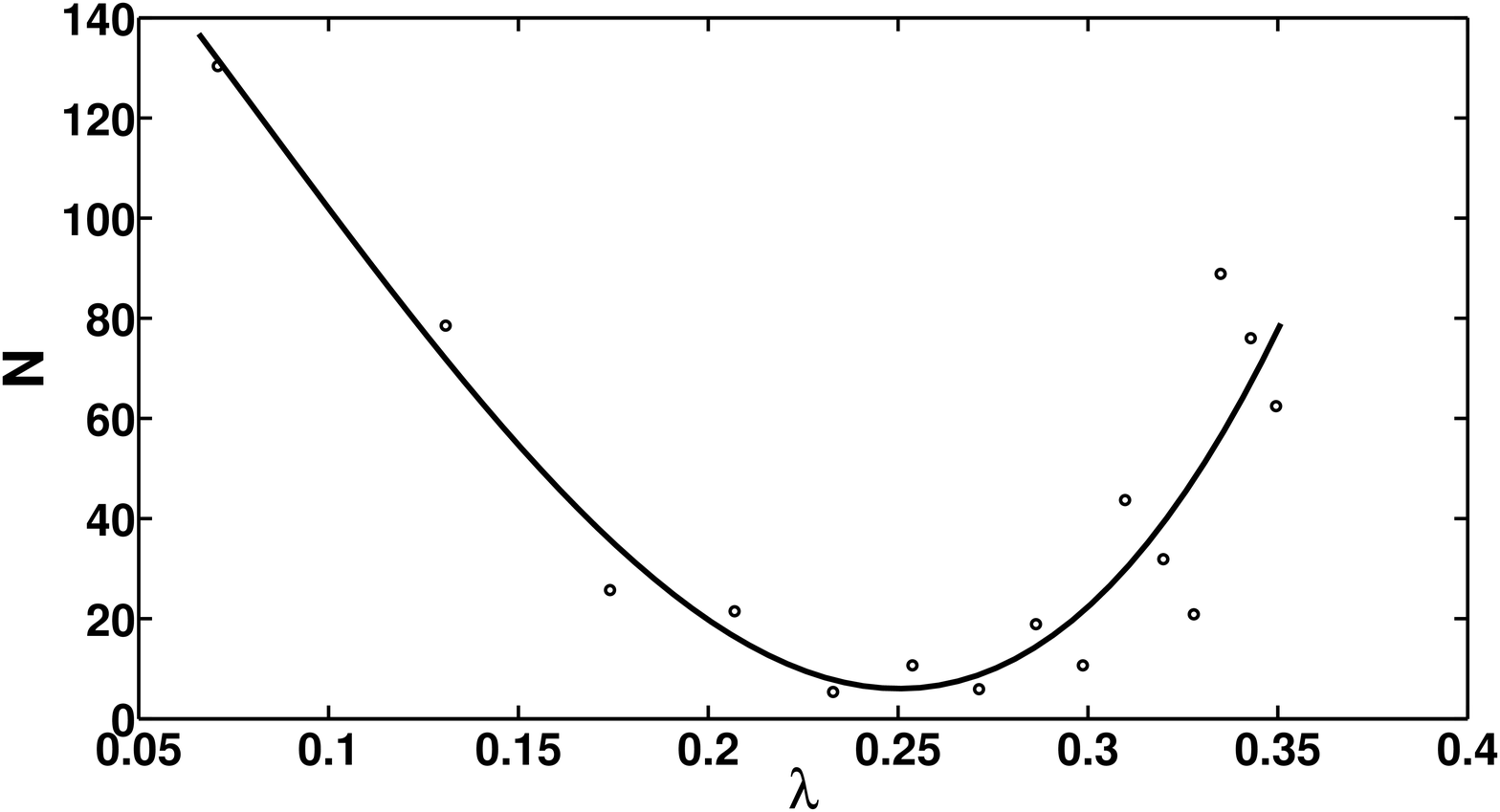}}
  \caption{Plasmon number versus the frequency shift for
  Eq.~\eqref{eq:NLSEpsi} in
  (a) 2D case and (b) 3D case. 
  \label{denseplBL}
  }
\end{figure}

Note that the general solution of Eq.~\eqref{ionosphBL} can be represented in the form $\psi = e^{\mathrm{i}\lambda\tau}\psi_0(x)$, where $\psi_0(x)$ is the new function. As in the case~(i), the form of $\psi_0$ is obtained numerically. In Fig.~\ref{denseplBL} we show the dependense $N(\lambda)$ for numerical solutions found in 2D and 3D cases.
It should be noted that the accuracy of calculations for 3D solitons is lower than in 2D case. Hence to build a smooth curve in Fig.~\ref{denseplBL}(b) the least squares method is used. Again applying VKC we get that 2D solitons are stable for a broad range of $\lambda$, and in 3D space both stable and unstable solutions are present.

We showed that in both (i) and (ii) cases there are stable spatial Langmuir solitons in 2 and 3 spatial dimensions. A 3D soliton is a spherically symmetric plasma structure which unambiguously models a natural BL. A soliton in the 2D space is an axially symmetric plasma oscillation and has a cylindrical form. A possible natural implementation of this kind of plasma structures can be a rare type of BL in the form of a glowing band~\cite{Byc10}.

In the case~(i) a soliton can be excited if the characteristic parameters of a background plasma are $n_0 = 10^{5}\thinspace\text{cm}^{-3}$ and $T_e = 10^6\thinspace\text{K}$. A plasma with such a density may well exist in the Earth ionosphere. The effective radius of a plasmoid in this case is $\sim 1\thinspace\text{m}$. Such a plasma structure resembles BLs observed near airplanes~\cite{Byc10}. In the case~(ii) we can discuss a plasma composed of nitrogen ions which are diatomic. Hence we adopt a reasonable model for the atmospheric plasma since it consists of mainly nitrogen. For the successful excitation of a soliton, the typical characteristics of plasma should be $n_0 = 10^{21}\thinspace\text{cm}^{-3}$, $T_e = 10^6\thinspace\text{K}$, and $T_i = 300\thinspace\text{K}$. The radius of a soliton is $\sim 10^{-5}\thinspace\text{cm}$, which is close to the prediction of the quantum model of BL~\cite{DvoDvo07}.

Note that the electron temperature used in our estimates is higher than that one can expect in a linear lightning bolt. Thus the developed model is likely to describe a protoBL, i.e. BL at the initial stages of its evolution. One can assume that under certain circumstances the characteristics of a background plasma can take the values favorable for the excitation of a Langmuir soliton within our model. At the later stages of its evolution other (maybe quantum) nonlinearities become important and a spatial soliton develops into a plasma object identified as BL.

It should be noted that, besides the description of BL, the model in the case~(ii) can be applied for the explanation of the results of experiments where long-lived glowing plasma structures were obtained in electric discharges in liquid nitrogen (see, e.g.,~\cite{KirSavKad95}). Indeed, if a plasmoid is excited in liquid nitrogen, we can take that $T_i = 77\thinspace\text{K}$, $T_e =10^4\thinspace\text{K}$, and $n_0 = 3.44 \times 10^{21}\thinspace\text{cm}^{-3}$. Such characteristics are well achievable in laboratory plasmas. It makes rather plausible interpretation of the results of~\cite{KirSavKad95} in frames of our model.

\textit{Acknowledgments} I am thankful to FAPESP (Brazil) for a grant.


\begin{thebibliography}{99}

\bibitem{Ara37}
  F.~Arago,
  \textit{Notices Scientifiques: Sur le Tonnerre} [in French]
  (Bachelier, Paris, 1837).

\bibitem{Byc10}\vspace{-8pt}
  V.~L.~Bychkov, \textit{et al.},
  in \textit{Atmosphere and Ionosphere: Dynamics, Processes and Monitoring},
  ed. by V.~L.~Bychkov, \textit{et al.}
  (Springer, Dordrecht, 2010), pp.~201--373.

\bibitem{DvoDvo07}\vspace{-8pt}
  M.~Dvornikov and S.~Dvornikov,
  in \textit{Advances in Plasma Physics Research, vol.~5},
  ed. by F.~Gerard
  (Nova Sci. Publ., New York, 2007), pp. 197--212
  [physics/0306157];
  M.~Dvornikov,
  Proc. R. Soc. A \textbf{468}, 415 (2012)
  [arXiv:1102.0944];
  J. Atm. Sol.-Terr. Phys. \textbf{89}, 62 (2012)
  [arXiv:1112.0239];
  J. Phys. A \textbf{46}, 045501 (2013)
  [arXiv:1208.2208].

\bibitem{Dvo11}\vspace{-8pt}
  M.~Dvornikov,
  J. Plasma Phys. \textbf{77}, 749 (2011)
  [arXiv:1010.0701].

\bibitem{Dvo13}\vspace{-8pt}
  M.~Dvornikov,
  Nonlin. Proc. Geophys. \textbf{20}, 581 (2013)
  [arXiv:1203.0258].

\bibitem{SkoHaa80}\vspace{-8pt}
  M.~M.~\v{S}kori\'{c} and D.~ter~Haar,
  Physica C \textbf{98}, 211 (1980);
  T.~A.~Davydova, \textit{et al.},
  Phys. Lett. A \textbf{336}, 46 (2005);
  G.~Simpson, \textit{et al.},
  Phys. Rev. E \textbf{80}, 056405 (2009).


\bibitem{KirSavKad95}\vspace{-8pt}
  D.~L.~Kirko, \textit{et al.},
  Tech. Phys. Lett. \textbf{21}, 388 (1995).

\end{thebibliography}
\end{document}